\documentclass[prl,twocolumn,superscriptaddress,floatfix,showpacs]{revtex4}

\usepackage{amsmath}
\usepackage[dvips]{graphicx}
\newcommand{\beq}{\begin{equation}}
\newcommand{\eeq}{\end{equation}}
\newcommand{\ket}[1]{\left\vert#1\right\rangle}
\newcommand{\bra}[1]{\left\langle#1\right\vert}
\newcommand{\Ham}{\mathcal H}

\begin{document}

\title{Effective thermal dynamics following a quantum quench in a spin chain}

\author{Davide Rossini}
\affiliation{International School for Advanced Studies (SISSA),
  Via Beirut 2-4, I-34014 Trieste, Italy}

\author{Alessandro Silva}
\affiliation{International Centre for Theoretical Physics (ICTP),
  I-34014 Trieste, Italy}

\author{Giuseppe Mussardo}
\affiliation{International School for Advanced Studies (SISSA),
  Via Beirut 2-4, I-34014 Trieste, Italy}
\affiliation{International Centre for Theoretical Physics (ICTP),
  I-34014 Trieste, Italy}
\affiliation{Istituto Nazionale di Fisica Nucleare, Sezione di Trieste, Trieste, Italy}

\author{Giuseppe E. Santoro}
\affiliation{International School for Advanced Studies (SISSA),
  Via Beirut 2-4, I-34014 Trieste, Italy}
\affiliation{International Centre for Theoretical Physics (ICTP),
  I-34014 Trieste, Italy}

\begin{abstract}

We study the nonequilibrium dynamics of the Quantum Ising Model
following an abrupt quench of the transverse field. We focus on the
onsite autocorrelation function of the order parameter,
and extract the phase coherence time $\tau^{\varphi}_Q$
from its asymptotic behavior.
We show that the initial state
determines $\tau^{\varphi}_Q$ only through an effective temperature
set by its energy and the final Hamiltonian. Moreover, we observe
that the dependence of $\tau^{\varphi}_Q$ on the
effective temperature fairly agrees with that obtained in thermal equilibrium
as a function of the equilibrium temperature.

\end{abstract}


%75.40.Gb   Dynamic properties (dynamic susceptibility, spin waves, spin diffusion, dynamic scaling, etc.)
%75.10.Jm   Quantized spin models
%73.43.Nq   Quantum phase transitions (see also 64.70.Tg Quantum phase transitions in equations of state, phase equilibria and phase transitions)
%03.65.Sq   Semiclassical theories and applications

\maketitle

A recent series of beautiful experiments with cold atomic
gases~\cite{bloch08,greiner02,kinoshita06} have triggered a great
deal of interest in some fundamental aspects of the non-equilibrium
dynamics of correlated quantum systems. The peculiarity of the
dynamics of cold atoms is its phase coherence on long time
scales. This was clearly demonstrated by the cycles of collapse and
revival of the order parameter observed in Ref.~\cite{greiner02}.
The interplay between phase coherence, strong interactions, and low
dimensionality may result in surprising effects: an example is the
lack of thermalization recently observed in quasi-one dimensional
condensates~\cite{kinoshita06}. The attribution of this phenomenon
to the closeness of these systems to integrability spurred an
intense discussion on the general relation between {\em quantum}
integrability and thermalization in the long-time dynamics of
strongly correlated quantum
systems~\cite{igloi00,sengupta04,calabrese06,rigol07,rigol08,kollath07,manmana07,Cramer08,barthel08,kollar08,Cazalilla06,gangardt08}.

The simplest nonequilibrium process to be considered in order to
study the long-time dynamics of a quantum system is the {\em quantum
quench}: an abrupt change in time of one of the system parameters or
of its boundary conditions. Recent studies of %quench dynamics in various
strongly correlated
models~\cite{igloi00,sengupta04,calabrese06,rigol07,rigol08,kollath07,manmana07,Cramer08,barthel08,kollar08,Cazalilla06,gangardt08}
have demonstrated that the behavior of integrable and non-integrable
systems can be quite different. Thermalization can be observed,
under specific circumstances, in nonintegrable
systems~\cite{rigol08,kollath07,manmana07}: asymptotic values of
significant observables, such as the momentum distribution function,
do not depend on the details of the initial state, but only on its
energy~\cite{rigol08}.
%The mechanism of such thermalization
%was conjectured to be analogous to the one proposed by
%Srednicki~\cite{Srednicki94} for systems with a classically chaotic
%counterpart.
On the other hand, for integrable systems thermalization does not
occur~\cite{sengupta04,calabrese06,rigol07,Cramer08,barthel08,kollar08,Cazalilla06,gangardt08}:
a larger amount of information on the initial state seems necessary
to predict the asymptotic state. It has been conjectured that this
information consists of the expectation value of a set of constants
of motion fixing in the Lagrange multipliers of a generalized Gibbs
ensemble~\cite{rigol07}. For a special quench in a 1D Bose-Hubbard
model~\cite{Cramer08} and for integrable systems with free
quasiparticles~\cite{barthel08}, the local reduced density matrix
was indeed proven to asymptotically tend to such generalized
ensemble. Moreover, the generalized Gibbs ensemble was shown to
correctly predict the asymptotic momentum distribution functions for
a variety of models and
quenches~\cite{calabrese06,rigol07,kollar08,Cazalilla06}. However,
it should be pointed out that neglection of correlations of the
occupation of different quasi-particle modes leads to incorrect
predictions for the noise and higher order
correlators~\cite{gangardt08}.

In this Letter, instead of focusing on the asymptotics of
observables, we take a different perspective, and study the
dependence on the initial state $\ket{\psi_0}$ of the intrinsic
time-scale of the dynamics after the quench. We do this by
considering the Quantum Ising chain, a prototypical
example of exactly solvable model with a quantum phase
transition~\cite{sachdev_book}. We study the autocorrelation
function of the order parameter after a quench of the transverse
field, extracting the phase coherence time $\tau^{\varphi}_Q$ from
its asymptotic exponential decay. We will show that, regardless of
the integrability of the model, the only information on the quench
needed to predict $\tau^{\varphi}_Q$ is the final gap $\Delta$ and
an {\em effective temperature} $T_{{\rm eff}}$, determined by the
energy of the initial state after the quench; moreover, the dependence of
$\tau^{\varphi}_Q$ on $T_{{\rm eff}}$ is in very good agreement with
that obtained, at equilibrium, for the same quantity
$\tau^{\varphi}_T$ as a function of the equilibrium temperature $T$.

The sharp contrast between the asymptotics of observables like the
transverse magnetization, determined by the entire set of
constants of motion, and the phase coherence time
$\tau^{\varphi}_Q$, depending just on $T_{\rm eff}$, has its
deep roots in the physics of the Quantum Ising
chain~\cite{sachdev_book}. This model, which can be diagonalized in
the continuum limit in terms of Majorana fermion quasi-particles
(see, e.g.~\cite{IZ}), possesses two sectors of operators~\cite{YZ}:
a local sector with respect to the quasiparticles, where the
$S$-matrix is simply $S=1$ and the model is equivalent to a free
theory, and a non-local sector, where $S=-1$ and the model describes
an interacting theory. In this respect, the non-local sector can be
used as low-energy theory of a more general class of models, not
necessarily integrable, belonging to the Ising universality class
(e.g., the non-integrable $\Phi^4$ Landau-Ginzburg model~\cite{GM}).
While the transverse magnetization belongs to the local sector,
implying sensitivity of its asymptotics to integrability, the
order parameter belongs to the nonlocal sector, making %the phase coherence time
$\tau^{\varphi}_Q$ representative of the Ising universality class.

We start by considering a spin-$1/2$ Quantum Ising chain in a %(dimensionless)
transverse magnetic field $\Gamma$ with periodic boundary conditions:
\beq
   \Ham(\Gamma) = -J \sum_j \Big[ \sigma^x_j \sigma^x_{j+1} + \Gamma \sigma^z_j \Big] \;,
   \label{eq:model}
\eeq
where $\sigma_j^\alpha$ ($\alpha = x,y,z$) are spin operators, $J$ is the interaction strength.
Hereafter, unless explicitly written, we set $J = 1$.
This system has a quantum critical point at $\Gamma_c=1$ separating two mutually dual gapped
phases, a quantum paramagnetic one ($\Gamma>1$) and a ferromagnetic one ($\Gamma<1$),
with energy gap $\Delta \equiv 2 \vert 1 - \Gamma \vert$.
At equilibrium, the presence of a critical point dramatically influences the temperature dependence
of the basic time-scale characterizing the system's dynamics: the phase coherence time
$\tau^{\varphi}_T$~\cite{sachdev_book}.
The latter is usually extracted from the asymptotics of the on-site spin autocorrelation function
$\rho^{xx}_T(t) \equiv \langle \sigma^x_j (t) \sigma^x_j (0) \rangle$,
which decays to zero exponentially,~\cite{Deift94,sachdev97}
$\rho^{xx}_T (t) \sim e^{-t/\tau^{\varphi}_T}$, at any finite temperature $T>0$, both at criticality ($\Delta=0$), and
in the off-critical region ($T \ll \Delta$).
At criticality~\cite{Deift94}, for $T \ll J$ one finds
%\beq
$   \tau^{\varphi}_T \simeq \frac{8}{\pi T}$,
%   \label{eq:tau_crit}
%\eeq
%
while $\tau^{\varphi}_T$ is exponentially larger~\cite{sachdev97,note} in the off-critical region with $T\ll \Delta$:
%
%\beq
$   \tau^{\varphi}_T \simeq \frac{\pi}{2 T} e^{\Delta/T}$.
%   \label{eq:tau_off}
%\eeq
%
%These formulas hold for very low temperatures, where infrared properties of the theory,
%and consequently finite-size effects, are important (see inset of Fig.~\ref{fig:Gamma_Ferro}).

%We will be interested in extracting a phase coherence time $\tau^{\varphi}_Q$ in the case of a quantum quench,
Consider now a quantum quench, which consists of preparing the system in the ground state corresponding
to a transverse field $\Gamma_0$, $\ket{\psi_0}=\ket{\psi(\Gamma_0)}$, and then abruptly quenching it, at $t=0$,
to some $\Gamma \neq \Gamma_0$.
For $t>0$, the state evolves unitarily under $\Ham(\Gamma)$, according to
$\ket{\psi_t} = \exp[-i \Ham(\Gamma) t] \ket{\psi(\Gamma_0)}$.
We define the zero-temperature on-site autocorrelation function describing the spin dynamics after the quench:
\beq
   \rho^{xx}_Q(t) \equiv \bra{\psi(\Gamma_0)} e^{i\Ham(\Gamma) t} \sigma^x_j e^{-i\Ham(\Gamma)t}
   \sigma^x_j \ket{\psi(\Gamma_0)} \;.
   \label{eq:XXcorr}
\eeq
%
%%%%%%%%%%%%%%%%%%%%%%%%%%%%%%%%%%%%%%%%%%%%%%%%%%%%%%%%%%%%%%%%%%%%%%%%%%%%%%%%%%%%%%%%%
\begin{figure}[!t]
  \begin{center}
    \includegraphics[scale=0.34]{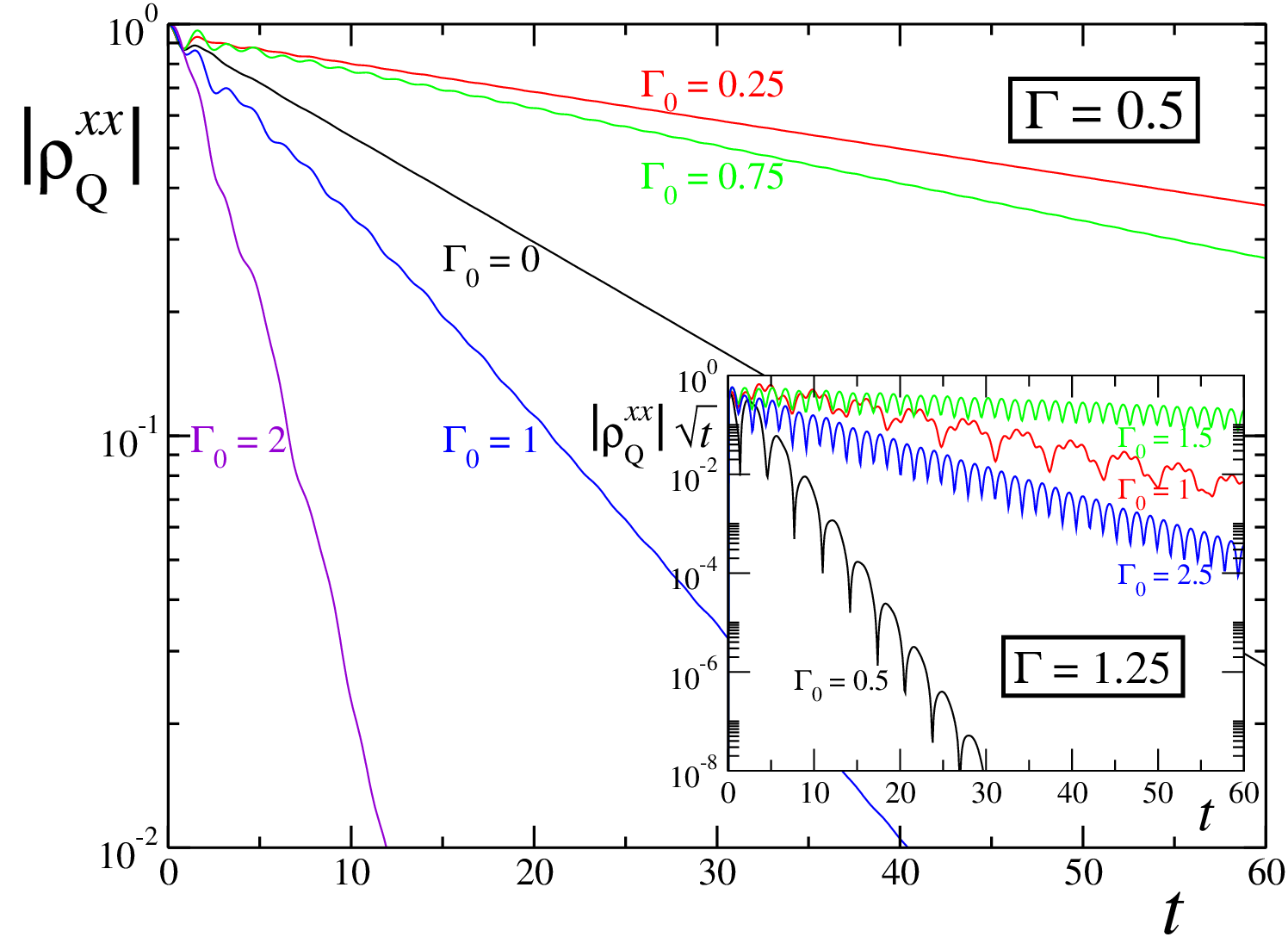}
    \caption{(color online).
      Time dependence of $|\rho^{xx}_Q(t)|$ for a quench to a final ferromagnet $\Gamma=0.5$.
      Different curves, obtained numerically for a finite chain of $L=600$ sites,
      correspond to different initial $\Gamma_0$'s.
      Inset: $|\rho^{xx}_Q(t)| \sqrt{t}$ for $\Gamma=1.25$.}% and different initial $\Gamma_0$'s.}
    \label{fig:CorrXX_Gfin050}
  \end{center}
\end{figure}
%%%%%%%%%%%%%%%%%%%%%%%%%%%%%%%%%%%%%%%%%%%%%%%%%%%%%%%%%%%%%%%%%%%%%%%%%%%%%%%%%%%%%%%%%
%
Before entering into details, we summarize the results obtained
by analyzing the asymptotics of $\rho^{xx}_Q(t)$: it always drops
exponentially to zero (see Fig.~\ref{fig:CorrXX_Gfin050}),
$\rho^{xx}_Q (t) \sim e^{-t/\tau^{\varphi}_Q}$, as in the
finite-temperature equilibrium case, consistent with what was obtained
in Ref.~\cite{calabrese06} for critical quenches. This allows us to
extract a time-scale $\tau^{\varphi}_Q$ characterizing the dynamics
after the quench.
This phase-coherence time depends in principle on
the initial state $\ket{\psi(\Gamma_0)}$
and the final Hamiltonian $\Ham(\Gamma)$.
However, and this is the main result of this Letter, all the
information needed to characterize $\tau^{\varphi}_Q$ is encoded in
two variables only: the final gap $\Delta(\Gamma)$, and an {\it
effective temperature} $T_{{\rm eff}}$. The latter is obtained by
comparing the energy associated to the initial state %$\ket{\psi(\Gamma_0)}$
with respect to the Hamiltonian %$\Ham(\Gamma)$
after the quench to the average energy of a {\it fictitious} thermal state
at temperature $T_{{\rm eff}}$ in an effective canonical ensemble:
\beq
   E(\Gamma_0)\equiv \bra{\psi(\Gamma_0)} \Ham(\Gamma) \ket{\psi(\Gamma_0)} = \langle \Ham(\Gamma) \rangle_{T_{{\rm eff}}} \,.
   \label{eq:Teff}
\eeq
Most importantly, we find that
$\tau^{\varphi}_Q=\tau^{\varphi}_{T=T_{\rm eff}}$, both for quenches
at criticality and away from it.
%, where $\tau^{\varphi}_T$ is the
%equilibrium phase-coherence time. % given by Eqs.~\eqref{eq:tau_crit}-\eqref{eq:tau_off}.

To calculate $\rho^{xx}_Q(t)$ we exploit
the complete integrability of the Ising chain~\cite{lieb61,pfeuty70}.
Here we sketch the essential steps~\cite{mccoy71}: first, one represents spins in terms of
Jordan-Wigner fermions
%
%\beq
$c_l \equiv \sigma^-_l \exp \big( i \pi \sum_{j=1}^{l-1} \sigma^+_j \sigma^-_j \big)$.
%   \label{eq:JWT}
%\eeq
%
Since the ground state has always an even number of fermions,
one can focus on the even $c$-fermionic Hilbert space sector.
Switching to momentum representation, the Hamiltonian is diagonalized
with a Bogoliubov rotation:
%
%\beq
$   \Ham(\Gamma) = \sum_{k>0} \epsilon_k^\Gamma \big( \gamma^\dagger_k \gamma_k +
   \gamma^\dagger_{-k} \gamma_{-k} -1 \big)$,
%\eeq
%
where $\gamma_k$ are fermionic quasi-particle operators,
$\epsilon_k^\Gamma = 2 \sqrt{\Gamma^2-2 \Gamma \cos k +1}$ is their dispersion,
and $k = \pm \frac{\pi (2n+1)}{L}$ with $n=0,\ldots, \frac{L}{2}-1$.
%($k \in (-\pi, \pi]$).
%
The second step consists in describing the dynamics after a quench.
This can be easily done in the Heisenberg picture~\cite{barouch70}, by solving the closed set of equations of motion
for the $c$-fermions in momentum space, with the initial conditions associated to the quench.
Finally, $\rho^{xx}_Q(t)$ is computed using a trick developed in Ref.~\cite{mccoy71}.
The operator $\sigma^x_j (t) \sigma^x_j (0)$ connects states with different $c$-fermion parity,
and it cannot be simply evaluated using Jordan-Wigner fermions in the even Hamiltonian sector.
This problem can be circumvented by considering a four-spin correlation function on
a chain of length $L$,
$C^{x} (t;L) = \big\langle \sigma^x_{1}(t) \: \sigma^x_1 (0)\:
               \sigma^x_{\frac{L}{2}+1}(t) \: \sigma^x_{\frac{L}{2}+1}(0) \big\rangle$.
This correlator conserves the c-fermion parity, and can be written as the square root of a
Pfaffian~\cite{mccoy71}, using the techniques of Ref.~\cite{lieb61}.
One finally recovers $\rho^{xx}_Q(t)$ using the cluster property $[\rho^{xx}_Q (t)]^2 = \lim_{L \to \infty} C^x(t;L)$,
by taking the square root of $C^{x}(t)$ in the limit of large number of spins.

As anticipated above, the zero-temperature quench autocorrelation
$\rho^{xx}_Q(t)$ always relaxes exponentially to zero (see Fig.~\ref{fig:CorrXX_Gfin050}),
irrespective of the initial state $\ket{\psi(\Gamma_0)}$ and of the final transverse field
$\Gamma \neq \Gamma_0$: $\rho^{xx}_Q (t) \sim e^{-t/\tau^{\varphi}_Q}$.
This is in sharp contrast with the zero-temperature equilibrium autocorrelation
$\rho^{xx}_{T=0}(t)$, which decays as $M_x^2 + C/t$ for $\Gamma<1$,
$M_x=(1-\Gamma^2)^{1/8}$ being the spontaneous magnetization~\cite{mccoy71}.
Quenching to the paramagnetic side, the exponential drop is
superimposed to an oscillatory power-law decay. This is once again
reminiscent of finite-temperature equilibrium case, where
$\rho^{xx}_T(t) \sim K(t) e^{-t/\tau^{\varphi}_T}$, $K(t)$ being the
quantum zero-temperature correlator~\cite{sachdev97}, which
oscillates and decays as $t^{-1/2}$.
Indeed, for a quench to $\Gamma>1$, rescaling $\rho^{xx}_Q(t)$ with
the zero-temperature factor $t^{-1/2}$, we recover
exponential relaxation  (inset of Fig.~\ref{fig:CorrXX_Gfin050}).

%%%%%%%%%%%%%%%%%%%%%%%%%%%%%%%%%%%%%%%%%%%%%%%%%%%%%%%%%%%%%%%%%%%%%%%%%%%%%%%%%%%%%%%%%
\begin{figure}[!t]
  \begin{center}
    \includegraphics[scale=0.34]{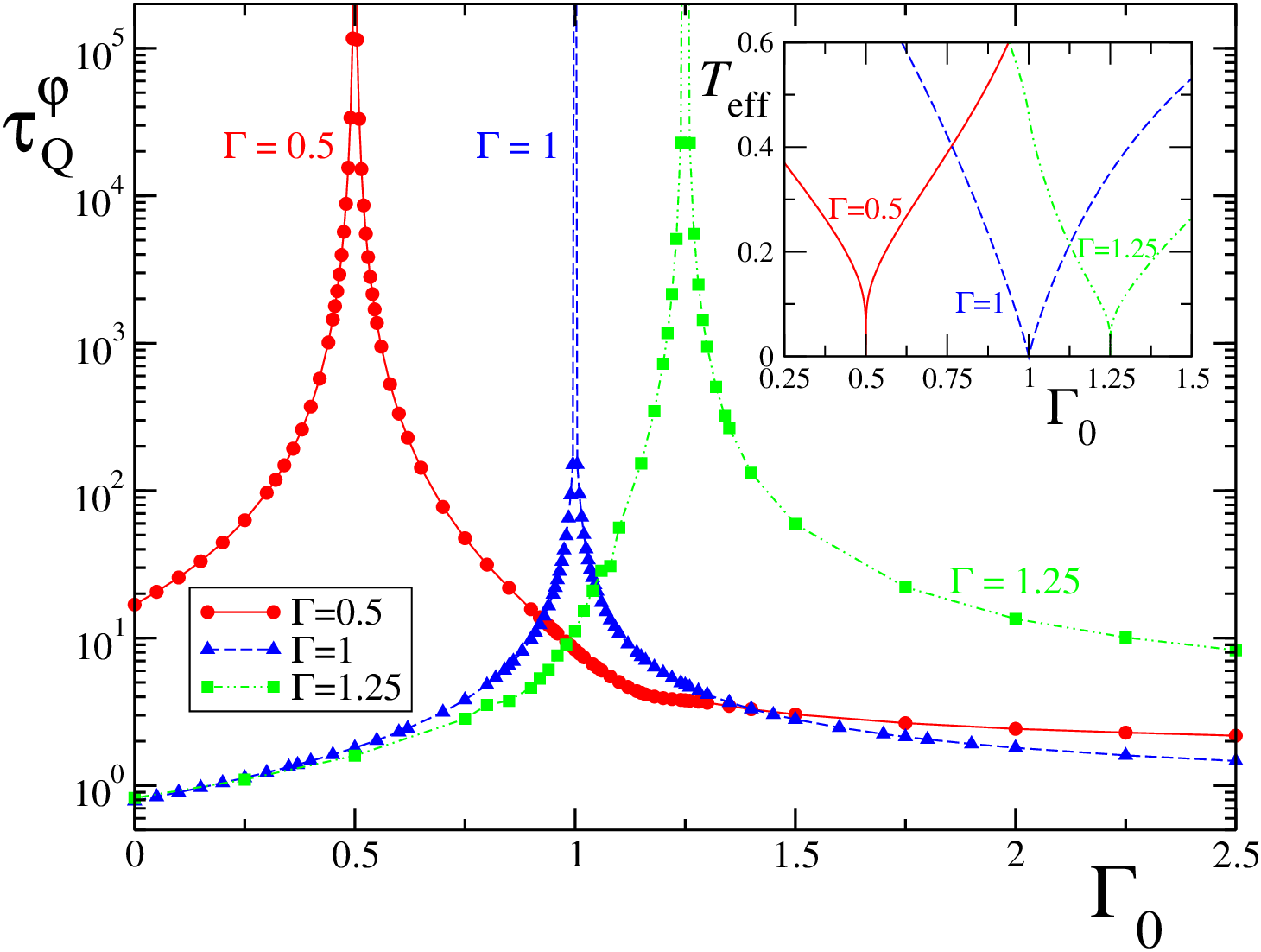}
    \caption{(color online). Phase coherence time $\tau^{\varphi}_Q$
      as a function of the initial transverse field $\Gamma_0$.
      The different curves refer to a ferromagnetic ($\Gamma = 0.5$, red circles),
      a critical ($\Gamma = 1$, blue triangles) and a paramagnetic
      ($\Gamma = 1.25$, green squares) quench dynamics.
      Inset: effective temperature $T_{\rm eff}$ vs. $\Gamma_0$, as extracted from Eq.~\eqref{eq:Teff},
      for the same values of $\Gamma$.}
\label{fig:Tau_gamma}
\end{center}
\end{figure}
%%%%%%%%%%%%%%%%%%%%%%%%%%%%%%%%%%%%%%%%%%%%%%%%%%%%%%%%%%%%%%%%%%%%%%%%%%%%%%%%%%%%%%%%%

%%%%%%%%%%%%%%%%%%%%%%%%%%%%%%%%%%%%%%%%%%%%%%%%%%%%%%%%%%%%%%%%%%%%%%%%%%%%%%%%%%%%%%%%%
\begin{figure}[!t]
  \begin{center}
    \includegraphics[scale=0.35]{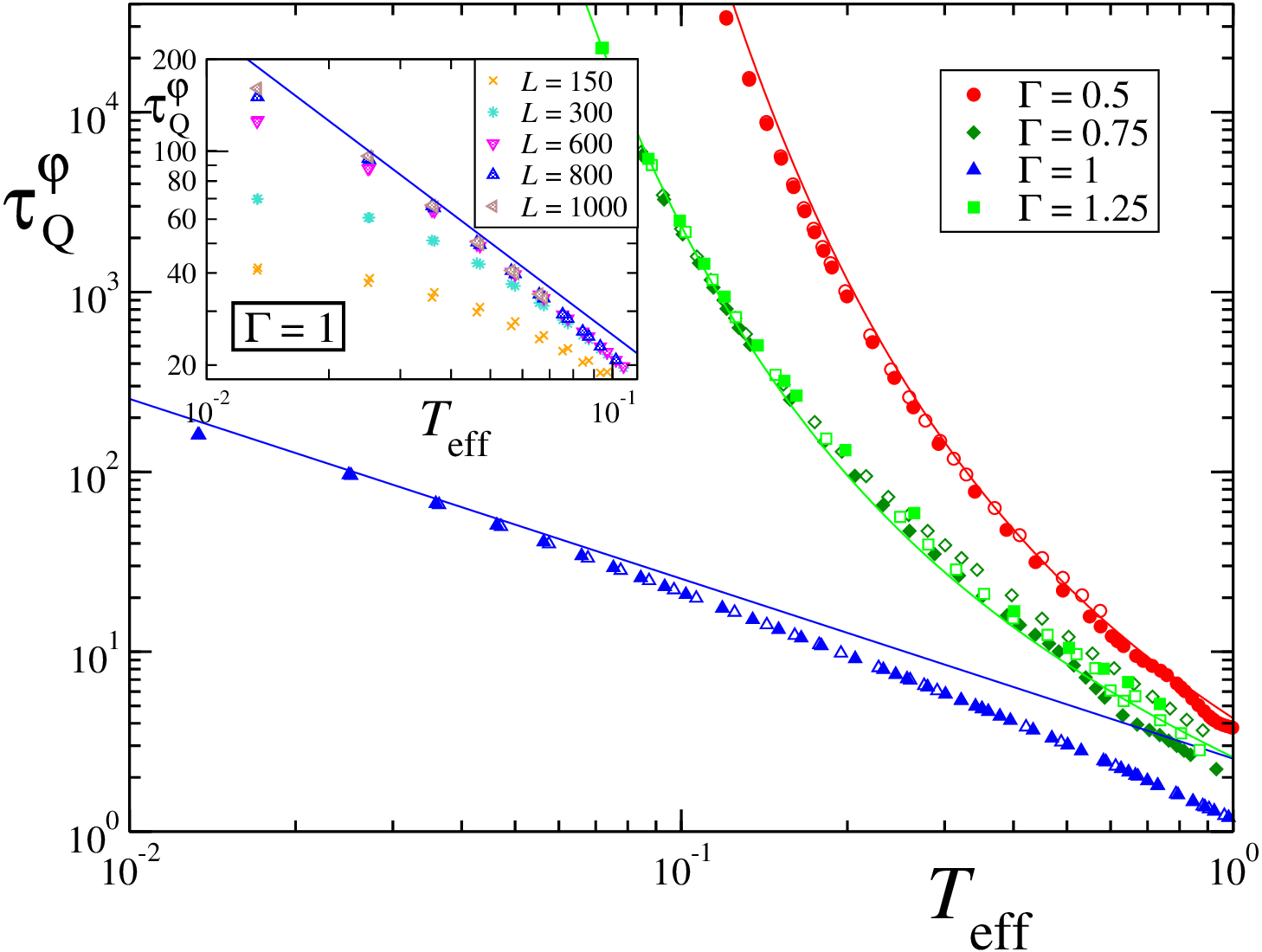}
    \caption{(color online). Phase coherence time $\tau^{\varphi}_Q$ as a function of $T_{\rm eff}$.
      Different symbols are for various values of $\Gamma$.
      Empty symbols correspond to an initial field $\Gamma_0 < \Gamma$,
      while filled ones are for $\Gamma_0 > \Gamma$.
      Straight lines denote the finite-temperature equilibrium values of $\tau^{\varphi}_{T=T_{\rm eff}}$.
      The inset shows the finite-size scaling of $\tau^{\varphi}_Q$ at criticality.
      Data in the main panel are for $L=600$
      (except for $\Gamma=1$ and $T_{\rm eff} < 0.07$, where $L = 1000$).}
    \label{fig:Gamma_Ferro}
  \end{center}
\end{figure}
%%%%%%%%%%%%%%%%%%%%%%%%%%%%%%%%%%%%%%%%%%%%%%%%%%%%%%%%%%%%%%%%%%%%%%%%%%%%%%%%%%%%%%%%%

We now analyze the coherence time as a
function of different initial and final conditions.
In Fig.~\ref{fig:Tau_gamma}, $\tau^{\varphi}_Q$ is plotted as a function of
initial $\Gamma_0$, for several final $\Gamma$'s.
A dramatic increase of $\tau^{\varphi}_Q$ as $\Gamma_0 \to \Gamma$ is
observed: the less the system goes out-of-equilibrium, the slower is
the relaxation. If $\Gamma_0=\Gamma$, the exponential
decay turns into a power-law, as in the zero-temperature equilibrium
case, and $\tau^{\varphi}_Q \to \infty$.
The analogy with the equilibrium finite-temperature behavior, where
the lower is the temperature $T$ the longer is $\tau^{\varphi}_T$,
is evident. It is therefore tempting to relate the two cases, by
introducing an effective temperature $T_{\rm eff}$ for the
out-of-equilibrium system. We define $T_{\rm eff}$ by comparing the energy
of the initial state $\ket{\psi(\Gamma_0)}$ with that of a
fictitious thermal state as in Eq.~(\ref{eq:Teff}), with a thermal energy
$\langle \Ham(\Gamma) \rangle_{T_{{\rm eff}}} = \sum_{k>0}
\epsilon_k^\Gamma (n_k(T_{\rm eff}) + n_{-k}(T_{\rm eff}) -1)$
determined by an effective canonical ensemble Fermi distribution function
$n_k(T_{\rm eff}) = (1 + e^{\epsilon_k^\Gamma/T_{\rm eff}})^{-1}$ of the
quasiparticles $\gamma_k$.
%
%
%($n_k = (1 + e^{\epsilon_k/T_{\rm eff}})^{-1}$)
%   \langle \psi_0 \vert \Ham \vert \psi_0 \rangle =
%   \sum_{k>0} \epsilon_k (n_k + n_{-k} -1) \,
%Here $n_k$ denotes the occupation number for the fermionic Bogoliubov
%quasiparticles $\gamma_k$ and is given, at the thermodynamic equilibrium,
%by the standard Fermi distribution function:
%$n_k = (1 + e^{\epsilon_k/T_{\rm eff}})^{-1}$.
%
A plot of $T_{\rm eff}$ as a function of $\Gamma_0$, for different
values of $\Gamma$, is shown in the inset of Fig.~\ref{fig:Tau_gamma}.
Notice that, for each $\Gamma$, there are {\em two} values
of $\Gamma_0$ for which $T_{\rm eff}$ is the same, one for
$\Gamma_0<\Gamma$ and one for $\Gamma_0>\Gamma$.

The effective temperature $T_{\rm eff}$, together with the
quasiparticle gap $\Delta$ at the final $\Gamma$, univocally
determines the phase coherence time $\tau^{\varphi}_Q$. Numerical
evidence is shown in Fig.~\ref{fig:Gamma_Ferro}: points with equal
$\Delta(\Gamma)$ have the same $\tau^{\varphi}_Q$ if effective
temperatures are the same, even if $\Gamma_0$ and $\Gamma$ are
different. Since the system is closed, it would be tempting to
substitute $T_{\rm eff}$ with just the initial energy $E(\Gamma_0)$.
This is not always possible. Indeed, two quenches having different
initial energy $E(\Gamma_0)$, but equal final gap $\Delta(\Gamma)$
and equal $T_{{\rm eff}}$, will exhibit the same $\tau^{\varphi}_Q$
(see data for $\Gamma=1.25$ and $0.75$ in
Fig.~\ref{fig:Gamma_Ferro}).
For example, the two quenches $1.21 \rightarrow 1.25$ and $0.715
\rightarrow 0.75$ have equal $T_{\rm eff} \simeq 0.113$ within
$0.7\%$, (corresponding to $\tau^{\varphi}_Q \simeq 1170$ within $0.3\%$),
although their energies differ by $25\%$.
However, if the final $\Gamma$ is fixed, the canonical and
microcanonical effective ensembles are equivalent, i.e.,
$E(\Gamma_0)$ and $T_{\rm eff}$ can be interchanged. We also notice
that, with a good accuracy, $\tau^{\varphi}_Q$ is still given by the
equilibrium expressions
%, Eqs.~\eqref{eq:tau_crit}-\eqref{eq:tau_off}),
% with the replacement $T \rightarrow T_{{\rm eff}}$,
at temperature $T_{{\rm eff}}$, i.e., $\tau^{\varphi}_Q\simeq
\tau^{\varphi}_{T=T_{\rm eff}}$ out of criticality. In the critical
case, a tendency to follow the equilibrium expressions
%(full curves in Fig.~\ref{fig:Gamma_Ferro})
is observed only at low temperatures,
using a finite-size scaling (see inset of
Fig.~\ref{fig:Gamma_Ferro}). This is necessary because, while the
comparison is expected to work better at low $T_{\rm eff}$, the
consequent importance of the long wavelength modes makes finite-size
effects more pronounced.
%
%A picture based on semiclassical arguments~\cite{sachdev97}
%justifying some of these findings is presented below.

%%%%%%%%%%%%%%%%%%%%%%%%%%%%%%%%%%%%%%%%%%%%%%%%%%%%%%%%%%%%%%%%%%%%%%%%%%%%%%%%%%%%%%%%%
\begin{figure}[!t]
  \begin{center}
    \includegraphics[scale=0.34]{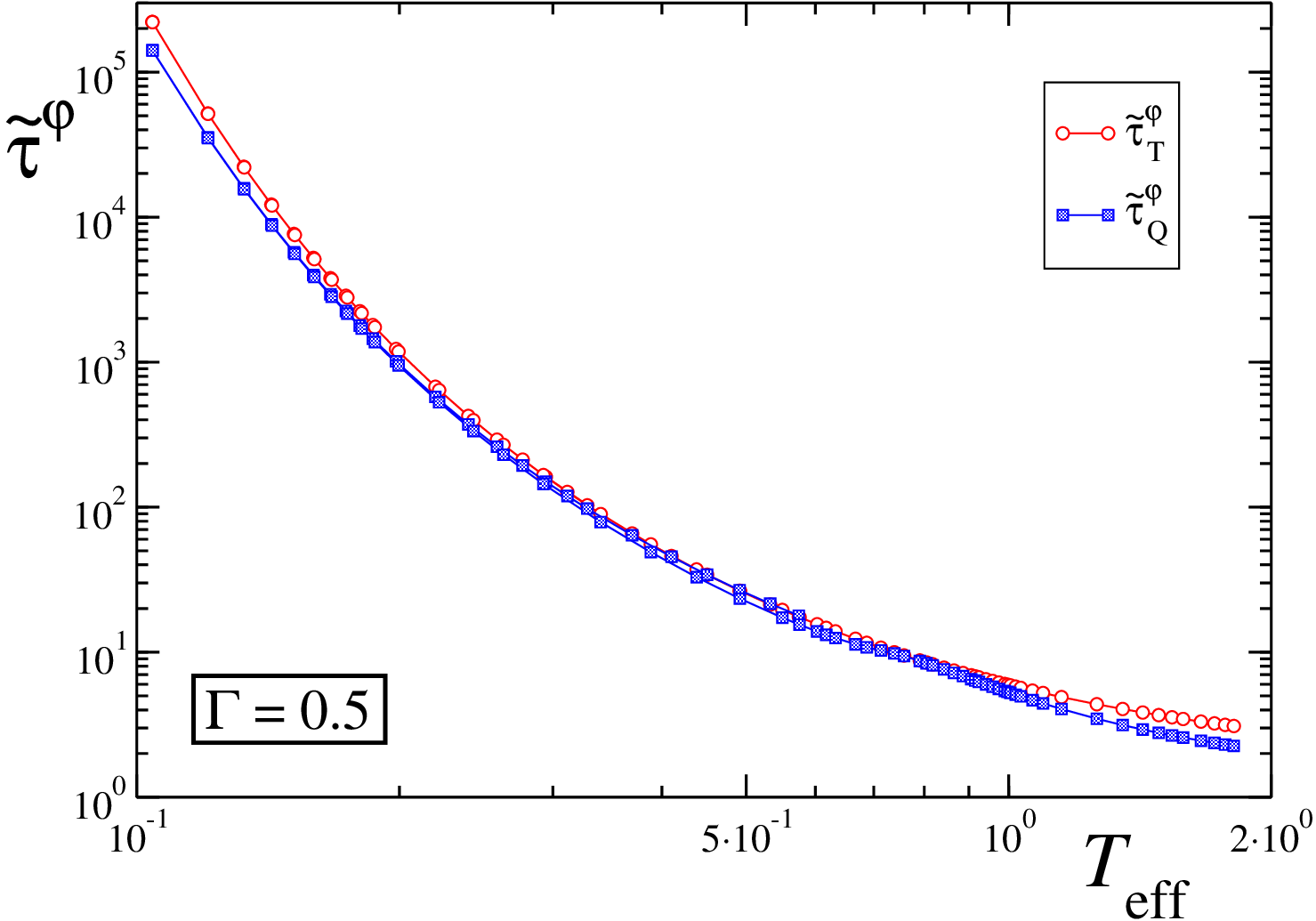}
    \caption{(color online). Phase coherence time at $\Gamma = 0.5$ computed according to the
      quasiparticle distribution $f_k$ ($\tilde{\tau}^{\varphi}_Q$, blue squares) and to the effective
      quasiparticle distribution $n_k(T_{\rm eff})$ ($\tilde{\tau}^{\varphi}_T$, red circles).}
    \label{fig:Temper_Gamma}
  \end{center}
\end{figure}
%%%%%%%%%%%%%%%%%%%%%%%%%%%%%%%%%%%%%%%%%%%%%%%%%%%%%%%%%%%%%%%%%%%%%%%%%%%%%%%%%%%%%%%%%

%..................... Semiclassical arguments .....................................................

The time dependence of correlators is influenced by quasiparticle
propagation, as exemplarily shown in a series of
studies~\cite{igloi00,sengupta04,calabrese06} focusing on the
asymptotics of \it equal time \rm correlators at different
space-points, $\rho^{xx}(r,t)=\langle \sigma^x_{i+r}(t)
\sigma^x_i(t) \rangle$.
%The most striking feature that emerged %in this context
%is the so-called light-cone
%effect~\cite{igloi00,calabrese06}: correlations of two spins at
%distance $r$ peak at time $t \simeq r/2v$, i.e., at the time when
%the first pair of quasiparticles reaches them, travelling at speed
%$v$ and originating from points placed between the two spins.
%
A similar picture, elucidating some of the results so
far obtained and applicable in the off-critical regimes ($T_{\rm eff} \ll \Delta$),
can be formulated in the spirit of Ref.~\cite{sachdev97}.
When the transverse field is quenched, the
initial condition for the time evolution consists of a state with a
finite density of quasiparticles (relative to $\Ham(\Gamma)$),
characterized by a dispersion $\epsilon_k^\Gamma$. For $\Gamma <1$,
these can be seen as kinks propagating with momentum
$k$ and velocity $v_k=\partial\epsilon_k^\Gamma/\partial k$.
The correlator $\rho^{xx}_Q(t)$ is determined by the number of
kinks passing through a single site in the interval $[0,t]$.
A combinatorial analysis~\cite{sachdev97}, together with an
average over momenta, leads to
$\rho^{xx}_Q(t) \simeq \exp[-t/\tilde{\tau}^{\varphi}_Q]$, with
%
%\beq
$   (\tilde{\tau}^{\varphi}_Q)^{-1} = \frac{2}{L} \sum_k \vert v_k \vert \, f_k$,
%   \label{tau}
%\eeq
%
where $f_k=\bra{\psi_0} \gamma^{\dagger}_k \gamma_k \ket{\psi_0}$ is
the occupation of quasiparticle modes. Analogous arguments can
be presented for $\Gamma>1$, giving
$\rho^{xx}_Q(t) \simeq K(t) \exp({-t/\tilde{\tau}^{\varphi}_Q})$,
where $K(t)\approx t^{-1/2}$ is the equilibrium zero-temperature correlator.
Note that here integrability is not necessary: the same reasoning
applies to the low-energy kink of non-integrable $\Phi^4$ theory that belongs to the
same Ising universality class.
While this picture explains the exponential decay towards zero,
it is also important to notice
the following: although the quasiparticle distribution function
$f_k$ determined by the initial state and the {\it effective}
thermal quasiparticle distribution function $n_k(T_{\rm eff})$ are
typically very different, phase coherence times $\tilde{\tau}^{\varphi}_Q$
computed from $f_k$ or according to
$(\tilde{\tau}^{\varphi}_T)^{-1} = \frac{2}{L} \sum_k \vert v_k \vert \, n_k(T_{\rm eff})$
are very close (see Fig.~\ref{fig:Temper_Gamma}).
In other words, we explicitly checked that, fixing an effective
temperature through Eq.~\eqref{eq:Teff} or using
$\tilde{\tau}^{\varphi}_Q=\tilde{\tau}^{\varphi}_T$, leads
to qualitatively and quantitatively similar (to a few percent accuracy)
results, for $\Delta \gg T_{\rm eff}$;
moreover, by imposing
$\tilde{\tau}^{\varphi}_Q=\tilde{\tau}^{\varphi}_T$, we find
$\tilde{T}_{\rm eff} \sim 2\Delta/\ln[\Delta/(\Gamma-\Gamma_0)^2]$
far from criticality and at low temperatures, in agreement with the cusp singularity of
Fig.~\ref{fig:Tau_gamma}.
%in the off-critical region ($\Delta \gg \Gamma$),
%As in Fig.~\ref{fig:Tau_gamma}, at criticality one obtains instead
%$T_{\rm eff} \approx {\rm const} \vert \Gamma -1 \vert$.

In conclusion, we studied the phase coherence time $\tau^{\varphi}_Q$ after an abrupt quench of
the transverse field in a Quantum Ising chain. We have shown that, irrespective of the
integrability of the model, $\tau^{\varphi}_Q$ depends only on %two variables,
the quasiparticle gap $\Delta$ and the quasiparticle effective
temperature $T_{\rm eff}$, and provided numerical evidence of the
fact that the dependence of $\tau^{\varphi}_Q$ on $T_{\rm eff}$ is
close to the one obtained at equilibrium as a function of the
equilibrium temperature. The realization of the dynamics of
artificial quantum spin chains using bosonic atoms in optical
lattices~\cite{Lukin} represents a concrete possibility to check our
theoretical scenario with the available experimental tools.

We thank N. Andrei, E. Altman, R. Fazio, V. Oganesyan,
V. Kravtsov, and A. Polkovnikov for discussions.
G.M. acknowledges hospitality at Galileo Galilei
Institute where this work was completed, and grants INSTANS
(from ESF) and 2007JHLPEZ (from MIUR).
%{\em Fisica Statistica dei Sistemi Fortemente Correlati all'Equilibrio e Fuori Equilibrio:
%Risultati Esatti e Metodi di Teoria dei Campi}

\vspace*{-2mm}
%---------------------------------------------------------------------------------------------------------

%---------------------------------------------------------------------------------------------------------


\begin{thebibliography}{99}
\bibitem{note}
  Given the bandwidth of the excitation $W=2J (\vert 1+\Gamma \vert
  - \vert 1-\Gamma \vert)$ we assume $T,T_{\rm eff} \ll W$.
\end{thebibliography}
\end{document}